\documentclass[prd, amsfonts, twocolumn, nofootinbib, showpacs]{revtex4}
\usepackage{graphicx, epsfig}
\usepackage{color}
\usepackage{amsmath}
\newcommand{\be}{\begin{equation}}
\newcommand{\ee}{\end{equation}}
\newcommand{\bea}{\begin{eqnarray}}
\newcommand{\eea}{\end{eqnarray}}

\newcommand{\gapp}{\mathrel{\raise.3ex\hbox{$>$}\mkern-14mu
\lower0.6ex\hbox{$\sim$}}}
\newcommand{\lapp}{\mathrel{\raise.3ex\hbox{$<$}\mkern-14mu
\lower0.6ex\hbox{$\sim$}}}
\def\bbox{{\,\lower0.9pt\vbox{\hrule \hbox{\vrule height 0.2 cm
\hskip 0.2 cm \vrule  height 0.2 cm}\hrule}\,}}

\begin{document}
\title{Green's function of a massless scalar field in curved space-time and superluminal phase velocity of the retarded potential}
\author{De-Chang Dai$^1$, Dejan Stojkovic$^2$}
\affiliation{ $^1$ Institute of Natural Sciences and INPAC, Department of Physics,
Shanghai Jiao Tong University, Shanghai 200240, China}
\affiliation{ $^2$ HEPCOS, Department of Physics, SUNY at Buffalo, Buffalo, NY 14260-1500}


\begin{abstract}
\widetext
We study a retarded potential solution of a massless scalar field in curved space-time. In a special ansatz for a particle at rest whose magnitude of the (scalar) charge is changing with time, we found an exact analytic solution. The solution indicates that the phase velocity of the retarded potential of a non-moving scalar charge is position dependent, and may easily be greater than the speed of light at a given point. In the case of the Schwarzschild space-time, at the horizon, the phase velocity becomes infinitely faster than the coordinate speed of light at that point. Superluminal phase velocity is relatively common phenomenon, with the the phase velocity of the massive Klein-Gordon field as the best known example. We discuss why it is possible to have modes with superluminal  phase velocity even for a massless field.
\end{abstract}


\maketitle
\section{Introduction}

The speed at which  fields (e.g. scalar, vector and gravitational) propagate is a very subtle question. According to the special relativity, energy (and mass) can not travel with speed greater than the speed of light, but there is nothing restricting the speed of auxiliary fields like potentials. Potentials describe interactions, and interactions are mediated by virtual particles, which are off-shell and do not have any a priori preferred speed.

For example, one is tempted to assume that the field of the non-moving source is frozen and does not propagate, until the source/charge moves and the field re-arranges its distribution. This would effectively mean that a static field is infinitely rigid and propagates with infinite speed. This may make sense classically, however, quantum mechanically interactions are fluctuations in space induced by virtual particles.  Therefore, the situation is dynamical. To find the speed at which some interaction propagates, one has to calculate explicitly the effects of retardation in the Green's function of the field that mediates that interaction.

It is well known that the QED vacuum structure can affect the propagation of light even in flat space. The so-called Scharnhorst effect is a phenomenon in which light signals travel faster in between the two closely spaced conducting plates, than outside of the plates \cite{Scharnhorst:1998ix}. The reason is the Casimir effect, i.e. the vacuum polarization effect is suppressed in between the plates, so the photon loses less time propagating in between the plates than outside. This gives a hint that a massless particles do not always propagate at the speed of light in vacuum.

There is even more counter-intuitive example in curved space-time. Namely, Drummond and Hathrell demonstrated in \cite{Drummond:1979pp}  that vacuum polarization is sensitive to the curvature of spacetime. For example, for a photon propagating in a curved space, vacuum polarization can induce
a modification of the wave equation in such a way so that in some cases photons travel at speeds greater
than unity.  The effect seems to be dispersive, and the phase velocity approaches the speed of light at high frequencies.  Since the high-frequency limit of the phase velocity determines causality, it seems like causality is preserved in case. Extensive discussion of this effect can be found in \cite{Hollowood:2010xh,Hollowood:2010bd,Hollowood:2008kq,Hollowood:2007ku,Hollowood:2007kt}.

These examples imply that propagation of quantum fields in curved space-time is a very subtle question, with many potential surprises.

The simplest and perhaps the most instructive case to study will be the case of the scalar field. The reason is that the scalar field potential is not gauge dependent. The only freedom we have is to add an extra constant, i.e. $\psi \rightarrow \psi +$const, which in turn has no dynamical effect and can be fixed by setting $\psi=0$ at infinity. The best way to find out the propagation velocity is perhaps to study the Green's function of a field. Once the Green's function is found, we can analyze the retarded potential for a given field and infer the speed at which the signal propagates from a point to a point. However, the difficulty of finding the general Green's function for a field in a curved space-time makes this approach very difficult. Fortunately, the full space Green's function is not absolutely necessary to study the propagation phenomena. A case in which an observer observes a modulated source will be sufficient to study. Therefore, we will consider a massless scalar field potential for a stationary (non-moving) but time-dependent source. Our result shows that the phase velocity of the retarded potentials is position dependent, and may easily be faster than the speed of light. In the case of the Schwarzschild space, this phase velocity at the horizon can even be infinitely greater than the speed of light at the horizon.  Though our solution in the linear ansatz is analytic, our analysis of the general form of the source is numerical. We therefore do not have an analytic form for a complete Green's function for an arbitrary source. However, the fact that the phase velocity of the scalar field varies locally  is important. Among the other things it implies that gravity must affect the path of the massless scalar field, which for example should lead to the gravitational lensing effect for the massless scalar field.

We would like to emphasis at the very beginning that throughout the paper we will use the term ``signal" in a loose sense. We will call any change in the field a ``signal". While a non-moving source does not emit any real particles, the phase of the field will change, and the speed of that change (phase velocity) we will call the speed of the signal. While this is not a real signal or information in a strict sense (i.e. transmitted by the group velocity) it will have some important consequences.

\section{Retarded Green's function for a massless scalar field in a curved space-time}

As a referent point, we first show the retarded Green's function for a massless scalar field in the flat space-time.
The speed at which the signal propagates through space can be read off the retarded solution. Consider a point particle in Minkowski space carrying a massless-scalar-field charge at the origin. Let the magnitude of its charge increase (or decrease) in time as $g(t)$. The equation of motion is
\begin{equation}
\label{motion1}
\partial_t^2 \psi -\partial_x^2\psi-\partial_y^2\psi-\partial_z^2\psi= 4\pi \delta (\vec{x})g(t)
\end{equation}
The solution for the function $\psi$ is
\begin{equation} \label{flat}
\psi=\frac{g(t-|\vec{r}|)}{|\vec{r}|} ,
\end{equation}
where $\vec{r}=(x,y,z)$. The scalar field potential falls off with distance in flat space as $1/r$. From the numerator, we see that the signal travels from a point to a point with the speed of light $c$, i.e., if we increase the magnitude of the charge at the origin, the potential at a point $\vec{r}$ will be affected after time $t=|\vec{r}|$. So, it takes some time for a signal to propagate even if the source is static. This is best understood in terms of virtual particles. A static source emits the sea of virtual particles which modify the space around it. Result in Eq.~(\ref{flat}) implies that, in flat space virtual (just like real massless) particles propagate with the speed of light, at least as long as they are in vacuum.

\section{Green's function for a massless scalar field in a curved space-time}
\label{sec:gf}

There are very few examples of exact Green functions in curved space-times \cite{Poisson:2003nc,Frolov:2012jj,Frolov:2003mc,Chu:2011ip,Chu:2012kz,Wiseman:2000rm,Burko:2002ge}. The reason is that it is notoriously difficult to find an exact solution without any approximations \cite{Bird:1975we,Ehlers:1976ji}. However, we will demonstrate that the particular case with a great degree of symmetry, i.e. a charge located at the center of a spherical symmetric curved space, is directly solvable. We will then use the explicit solution to discuss the speed at which scalar field potentials propagate in such space-times.

We fix again a point scalar charge at the origin of a spherically symmetric space. Fixing the charge at the origin rather than at an arbitrary point in space will provide the required symmetry and greatly facilitate the problem. We let its magnitude change as $g(t)$. The geometry of the space-time can be written as
\begin{equation}
d\tau ^2 =g_{tt}dt^2+g_{rr} dr^2-r^2 \left(d\theta^2 +\sin^2\theta d\phi^2 \right)
\end{equation}
The equation that we will try to solve is
\begin{equation}
\label{motion2}
D^tD_t\psi +D^rD_r\psi+D^\theta D_\theta\psi+D^\phi D_\phi\psi= \delta (r)\frac{g(t)}{\sqrt{h}}
\end{equation}
where $h=-g_{tt}g_{rr}r^4$.

We note here that our definition of the scalar charge slightly differs from the definition in Eq.~2.2 in \cite{Wiseman:2000rm}. The difference is the time component of the four-velocity $u^t$, which in our static case (charge is not moving) is just a constant and can be absorbed in $g(t)$. Further, strictly speaking, we are dealing with geometries  without horizons, so we will not discuss the no-hair theorems.

We will first try to find the time-independent solution, which will correspond to a static scalar charge of constant magnitude. In that case, Eq.~(\ref{motion2}) reduces to

\begin{equation}
\frac{1}{\sqrt{h}}\partial_r \Big( g^{rr}\sqrt{h}\partial_r \psi\Big)=\delta (r)\frac{g(t)}{\sqrt{h}}
\end{equation}

Without loss of generality, we can set $g(t) = 1$. The solution can be found by applying volume integration over the element $\sqrt{h}dtd^3 x$. The static solution $\psi_s$ is
\begin{equation}
\label{state1}
\psi_s(r)=\int_r^{\infty}  \sqrt{\frac{-g_{rr}(R)}{g_{tt}(R)}}\frac{1}{R^2} dR
\end{equation}
It is easy to verify that this is a solution by substituting Eq.~(\ref{state1}) back into Eq.~(\ref{motion2}).
Since this solution depends on both $g_{tt}$ and $g_{rr}$, it is different from the result from the flat space, but it will reduce to the flat space solution at large radius $r$.

We will now try to find the time-dependent solution which corresponds to $g(t)$.
Since we a priory expect massless particle to propagate with the speed of light, we may expect this solution to have the following form
\begin{equation}
\psi (t,r)= g\left( t-\int_0^{r} \sqrt{\frac{-g_{rr}(R)}{g_{tt}(R)}}dR \right)\psi_s(r) ,
\end{equation}
where the expression in parentheses on the right-hand side is the argument of the function $g$. This form is a straightforward curved space generalization of the flat space solution given by Eq.~(\ref{flat}).
The term $\sqrt{\frac{-g_{tt}}{g_{rr}}}$ is just the coordinate speed of light in the radial direction (obtained from $d\tau =0$). Though this appears to be a reasonable guess, this form is a solution only if $g_{rr}=-g_{tt}$. However, this requirement brings us back to the flat space. We will therefore generalize the form of the solution allowing for the possibility that the propagation speed is not the speed of light. We now try a more general form
\begin{equation}
\psi (t,r) = g\left( t-\int_0^{r} \frac{1}{v(R)}dR\right)\psi_s(r)
\end{equation}
where $v(r)$ is the coordinate speed at which the signal of the retarded potential travels (not necessarily the speed of light). This $v(r)$ must asymptotically go into the speed of light at large $r$ where the space-time becomes flat. We do not expect this form to always generate a solution.
However, if $g$ is a linear function of its argument, one can find the suitable solution. Therefore we consider the linear form of $g$
\begin{equation}\label{la}
g\left( t-\int_0^{r} \frac{1}{v(R)}dR\right)=A \left( t-\int_0^{r} \frac{1}{v(R)}dR\right)+B
\end{equation}
where $A$ and $B$ are two constants. If we plug this ansatz into Eq.~(\ref{motion2}), we find a condition under which the solution is valid
\begin{equation} \label{v}
v=\sqrt{\frac{-g_{tt}}{g_{rr}}}r^2 \psi_s^2= r^2 \psi_s^2c_l
\end{equation}
Here, $c_l=\sqrt{\frac{-g_{tt}}{g_{rr}}}$ is the coordinate speed of light. Since $v$ is also a coordinate velocity of propagation, it will be different for different observers and it will change from point to point. But, it is clear that the retarded signal propagates at a speed that is different from the speed of light for a given observer.

To make this more clear, we consider the Schwarzschild space-time, i.e.
\begin{equation}
g_{tt}=-g_{rr}^{-1}=1-\frac{2m}{r}
\end{equation}
We plug this condition into Eq.~(\ref{state1}), and the static solution becomes
\begin{equation}
\psi_s(r)=\int_r^{\infty} \frac{1}{1-\frac{2m}{R}} \frac{1}{R^2} dR=\frac{-\ln(1-\frac{2m}{r})}{2m} \neq \frac{1}{r}
\end{equation}
We see that the static scalar field potential does not fall off as $1/r$ which was the case in flat space. However, in the limit of $r \gg 2m$, we recover the usual $1/r$ behavior.

From Eq.~(\ref{v}), the coordinate propagation speed of the retarded potential is
\begin{equation} \label{vphase}
v= r^2 \psi_s^2c_l>c_l
\end{equation}
which is not equal to the coordinate speed of light $c_l$, and in fact is always greater than $c_l$ (for this particular example of the Schwarzschild space-time).
Our calculations will be strictly valid as long as our space-time is not strictly a black hole, but the conclusions will be valid even when we are only slightly outside the horizon.
In the extreme limit, exactly at the horizon, $r=2m$, the propagation speed, $v$, becomes infinitely faster than the speed of light at that point.

In the context of the Schwarzschild black hole, the coordinate speed of light, $c_l$, vanishes at the horizon, and any signal sent from the horizon gets infinitely redshifted. Thus, in the standard description it remains unclear how information about the black hole charge which is presumably imprinted at the horizon can be communicated to the region around the black hole.
[If the scalar charge is conserved, then formation of the black hole can not violate this conservation \cite{Stojkovic:2005zq}.]
Let's check what happens when the propagation speed of the retarded potential is taken into account. We can calculate the time, $\Delta t$, for a signal to propagate from the horizon to some finite distance $R$
\begin{eqnarray}
\Delta t &=&\int_{2m}^R \frac{1}{v} dr\nonumber\\
&=&\int_{2m}^R \frac{1}{1-\frac{2m}{r}} \ \frac{dr}{\left[r\ln(1-\frac{2m}{r})\right]^2} \nonumber\\
&= &\frac{-2m}{\ln(1-2m/R)}
\end{eqnarray}
This time is finite and therefore the potential has no problem to propagate from the horizon outside. Thus, a charged particle (at least with the scalar charge) can keep communicating its potential to the region outside the black hole. We mention again that our results are not strictly applicable to the black hole case, but we can always consider a shell whose radius is just slightly outside its own Schwarzschild radius and preserve the qualitative conclusions drawn here. In the next section we will reveal that the propagating velocity, $v$, is the phase velocity. Thus, this velocity refers to the change of phase, and it is not a group velocity. The casual light cone for real particles remains the same,  the Green's function does not have support outside the light cone, and causality is preserved.

\section{General case}

The discussion of the time dependent source so far was based on a solution found in the particular ansatz of Eq.~(\ref{la}). We will now try to analyze the general form of $g(t)$ (not only the linear ansatz that we used). We will first decompose the source $g(t)$ into different frequency modes as
\begin{equation}
g(t)=\int \tilde{g}(\omega)\exp(i\omega t) d\omega
\end{equation}
The wave number of each frequency mode is $\omega/v_\omega (r)$, where $v_\omega$ is the phase velocity of that mode. Then, the scalar potential can be written as
\begin{equation}
\psi(t,r) = \int \bar{g}(\omega)\exp\left[i\omega (t-\int \frac{1}{v_\omega}  dr)\right]f_\omega(r) d\omega
\end{equation}
where $f_\omega$ is the amplitude of the mode labeled by the frequency $\omega$. $\bar{g}(\omega)$ and $\tilde{g}$ can be found by matching the boundary condition at $r=0$.
By plugging the above equation into equation (\ref{motion2}), we find that $f_\omega$ and $v_\omega$ must satisfy
\begin{eqnarray}
\label{wave4}
&&-\frac{\omega^2}{g_{tt}}f_\omega -\frac{\omega^2}{v_\omega^2 g_{rr}}f_\omega +\frac{1}{\sqrt{h}}\partial_r \Big( \frac{\sqrt{h}}{g_{rr}}\partial_r f_\omega\Big)\nonumber\\
&&-i\omega\Big(\frac{\partial_r f_\omega}{v_\omega g_{rr}}+\frac{1}{\sqrt{h}}\partial_r (\frac{\sqrt{h}}{g_{rr}v_\omega}f_\omega)\Big)=0
\end{eqnarray}
except at $r=0$.
Since both imaginary and real parts must vanish independently, the above equation can be rewritten as two equations
\begin{eqnarray}
\label{wave5}
&&-\frac{\omega^2}{g_{tt}}f_\omega -\frac{\omega^2}{v_\omega^2 g_{rr}}f_\omega +\frac{1}{\sqrt{h}}\partial_r \Big( \frac{\sqrt{h}}{g_{rr}}\partial_r f_\omega\Big)=0\\
\label{wave-v}
&&\frac{\partial_r f_\omega}{v_\omega g_{rr}}+\frac{1}{\sqrt{h}}\partial_r \Big(\frac{\sqrt{h}}{g_{rr}v_\omega}f_\omega\Big)=0
\end{eqnarray}

Eq.~(\ref{wave-v}) can be easily solved by integrating with respect to $r$.

\begin{equation} \label{eq}
f_\omega^2\frac{\sqrt{h}}{g_{rr}v_\omega}= {\rm constant}
\end{equation}
In the limit of $r\rightarrow \infty$, the space becomes flat, which implies $v_\omega \rightarrow 1$ and $f_\omega \rightarrow 1/r$, as it should.

Eq.~(\ref{eq}) can be rewritten as
\begin{equation}
\label{velocity1}
v_\omega=P_\omega^2\sqrt{\frac{-g_{tt}}{g_{rr}}}
\end{equation}
where $P_\omega \equiv f_\omega r$.
If we substitute this relation into Eq.~(\ref{wave5}), and replace $f_\omega r$ with $P_\omega$ we get

\begin{eqnarray} \label{veleq}
&&\partial_r^2 P_\omega+\frac{1}{2}\partial_r\ln \left(\frac{-g_{tt}}{g_{rr}}\right)\partial_rP_\omega-\frac{1}{2}\partial_r\ln \left(\frac{-g_{tt}}{g_{rr}}\right)\frac{P_\omega}{r}\nonumber\\
\label{wave-6}
&&-\frac{\omega^2 g_{rr}}{g_{tt}}\left(P_\omega -\frac{1}{ P_\omega^3}\right) =0
\end{eqnarray}

The zero mode, $\omega =0$, solution to this equation is exactly $f_\omega =\psi_s$, where $\psi_s$ is the time-independent $g=$const solution given in Eq.~(\ref{state1}). Moreover, the phase velocity $v_\omega$  in this case is the same as the propagation velocity given by Eq.~(\ref{v}) in the ansatz solution we found. This then reveals the meaning of the parameter $v$ in Sec. \ref{sec:gf}.

In the high frequency limit, $\omega \rightarrow \infty$, the last term in Eq.~(\ref{veleq}) dominates. In order to satisfy the equation it has to vanish, thus requiring $P_\omega  =1$. Eq.~(\ref{velocity1}) then implies that $v_\omega $ becomes the speed of light $c_l$. It is this feature that ensures causality.

For the $\omega \neq 0$ modes, we will again use the spherically symmetric Schwarzschild geometry. The boundary conditions are
\begin{eqnarray}
\mbox {$r\rightarrow \infty$, $P_\omega =1$ }\\
\mbox {$r\rightarrow \infty$, $\partial_r P_\omega =0$ }\\
\mbox {$\frac{-g_{tt}}{g_{rr}}=(1-1/r)^2$}
\end{eqnarray}

\begin{figure}[h]
  \centering
\includegraphics[width=3.2in]{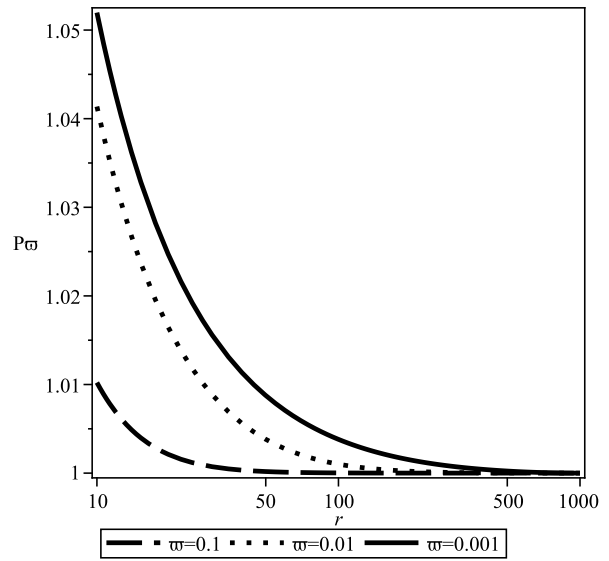}
\caption{This figure shows $P_\omega(r)$ for three different values of $\omega$, i.e. $\omega=10^{-1}$, $\omega=10^{-2}$ and $\omega =10^{-3}$. We see that $P_\omega$ grows as it is approaching the origin, higher $\omega$ modes increase slower than lower $\omega$ modes, and $P_\omega \geq 1$ everywhere. This behavior implies that the phase velocity $v_\omega$ is always superluminal (for this case of the Schwarzschild geometry), and that lower $\omega$ modes propagate faster than the higher $\omega$ modes.}
    \label{velocity}
\end{figure}

 In Fig.~\ref{velocity}, we show $P_\omega$ as a function of $r$ for several values of the frequency $\omega$. We can see that $P_\omega$ grows as it is approaching the origin. It is also apparent that higher $\omega$ modes increase slower than lower $\omega$ modes. We do not plot $P_\omega$ near the horizon, because the singularity will cause numerical instabilities. Since $P_\omega \geq 1$ for all values of $r$, the phase velocity defined by Eq.~(\ref{velocity1})
\begin{equation}
v_\omega=P_\omega^2\sqrt{\frac{-g_{tt}}{g_{rr}}}\geq c_l
\end{equation}
is greater than the speed of light everywhere.

\section{Common features with other examples with superluminal phase velocity}

In this section we discuss some other known examples where the phase velocity is superluminal, which may have have something in common with our results.

Perhaps the best known example is that of a massive Klein-Gordon field in flat space-time. The dispersion relation is simply $\omega^2=k^2+m^2$.
The phase velocity
\be
v_p \equiv \omega/k
\ee
is always greater than unity as long as $m \neq 0$. Moreover, superluminality is most pronounced for low frequencies $\omega$, while for large frequencies we have $\omega \approx k$, i.e. $v_p \approx 1$. However, the group velocity
\be
v_g \equiv d\omega/dk
\ee
is always less than unity. Comparing this result with the results we obtained in curved space, we might conclude that the curvature of space induces an effective mass to the massless scalar field, making it formally equivalent to the massive Klein-Gordon field with superluminal phase velocity.

The other, less known example is a massless scalar field in a $(5+1)$-dimensional flat space-time. The wave equation is
\begin{equation}
\partial_t^2 \psi -\partial_{x_1}^2\psi-\partial_{x_2}^2\psi-\partial_{x_3}^2\psi-\partial_{x_4}^2\psi-\partial_{x_5}^2\psi= \delta (\vec{x})\delta(t)
\end{equation}
The general solution can be found in most mathematical physics textbooks (e.g. \cite{Hassani}) or papers \cite{Cardoso:2002pa}. The Green's function for this case is
\begin{equation}
G^{5+1}(t,r)=-\frac{1}{8\pi^2}\Big(\frac{\delta'(t-r)}{r^2}+\frac{\delta(t-r)}{r^3}\Big)
\end{equation}
where, $r=\sqrt{x_1^2+x_2^2+x_3^2+x_4^2+x_5^2}$. If one considers the following concrete source
\begin{equation}
f(t,\vec{x}) =\sin(\omega t)\delta(\vec{x})
\end{equation}
then the wave function can be easily found as
\begin{eqnarray}
\psi&=&\int f(t',\vec{x'}) G^{5+1}(t-t',\vec{x}-\vec{x'})dt' d\vec{x'}\\
&=&-\frac{1}{8\pi^2}\Big( \frac{\omega \cos(\omega (t-r))}{r^2}+\frac{\sin(\omega (t-r))}{r^3} \Big)
\end{eqnarray}
Since this form includes two trigonometric functions, it is hard to see how the phase changes. We will then combine the two terms into a single trigonometric function.
\begin{eqnarray}
\psi&=&S(\omega,r)\sin(\omega(t-R))\\
S(\omega,r)&=&-\frac{1}{8\pi^2}\frac{\sqrt{1+r^2\omega^2}}{r^3}\\
R&=&r-\phi(r)/\omega\\
\phi(r)&=&\sin^{-1}\Big(\frac{r\omega}{\sqrt{1+r^2\omega^2}}\Big)
\end{eqnarray}

This form is similar to the form we studied in the last section. The phase velocity in this case is
\begin{equation}
v_p =\frac{1}{\partial_r R}=1+\frac{1}{r^2\omega^2}
\end{equation}

We can see that the phase velocity is infinite at the origin (for fixed $\omega$) but equal to the speed of light at $r \rightarrow \infty$. In this case the solution is created by two waves with different phases. Since their amplitudes decay in different ways, their combination makes the total phase velocity change with location, and in fact makes it infinitely faster than the speed of light at some locations. These are the features which are shared with our solution for the massless scalar field in a curved space. Moreover, for a fixed finite $r$, superluminality is again most pronounced for small $\omega$, while for large frequencies we have $v_p \approx 1$.


\section{Conclusions}

In this paper we analyzed the question of the speed at which potentials propagate in curved space-time. While finding an answer is easy in flat space, it becomes highly non-trivial in curved space-time. The difficulties range from finding an exact solution for the Green's function to choosing the right definition of the propagation speed. To avoid gauge and other ambiguities we considered the massless scalar potential. We located the scalar charge whose magnitude was changing in time at the origin in a spherically symmetric space-time, and found the solution for different frequency modes for this configuration. A non-moving particle does not emit real scalar field quanta, but what is changing in the system is the phase of the field. We found that the phase velocity is not constant but changes from point to point. Moreover, in the specific case of the Schwarzschild geometry, it is always greater than the coordinate speed of light at any given point. In an extreme limit, exactly at the horizon, the phase velocity becomes infinity faster than the speed of light at that point (which is actually vanishing). In fact, this feature is required if a black hole is going to communicate "information" about its potential which is presumably located at the horizon to the outer world.

It is important to note that the phase velocities, $v_\omega$, for different frequency modes (labeled by the frequency $\omega$) are different for each mode, and in general they are different from the local speed of light. Also, the amplitudes of different frequency mode's ($f_\omega$ in the text), have different $r$ dependence. These two facts make the curved space case quite different from the flat space, and explain why it was impossible to find a uniform propagation mode like in flat space (see Eq.~(\ref{flat}).
Since the propagation is dispersive, the high-frequency limit of the phase velocity will determine causality. Since the phase velocity approaches the speed of light at high frequencies, causality is preserved in our case.

The cases of the electromagnetic and gravitational potentials are more complicated because of the non-zero spin. However, they are also massless fields and will perhaps have some similar properties. In particular, we expect the retarded electromagnetic and gravitational potential from a non-moving source to propagate at a speed different from the speed of light. In other words, the average velocities of virtual photons and gravitons should not be the same as for real photons and gravitons in curved space-times.

A related question can be asked in the context of gravitational lensing. If the retarded gravitational potential of a static source travels with a finite speed (not necessarily the speed of light), it must experience the effect of the gravitational lensing, just as the light does. This would imply that the gravitational lensing effect on gravitons should be able to amplify or reduce the strength of gravity from a given static source \cite{Nemiroff:2005az}. In \cite{Wucknitz}, several examples were constructed to emphasis that the gravitational lensing could affect real gravitons, but could not lens any static gravitational field potential (though the static potential could be affected to some extent). However, the sources used in these examples were infinite planes, and not point sources, so the conclusions are perhaps not general.

If our conclusions for the scalar field hold for gravitons as well, then the static gravitational potential could propagate at any finite speed (except in the extreme case of the black hole horizon where it should be infinite) depending on the curved background. Since this speed is finite and position dependent, the effect of gravitational lensing of gravity should exist, though the magnitude of the effect should be different from the gravitational lensing of the light because of the different speed of propagation.
It is interesting that one of the possible explanations of the Pioneer anomaly \cite{Nieto:2003rq} is the focusing of gravity at around $25$AU \cite{Nemiroff:2005az}, exactly where the Pioneer anomaly arises. This could be a hint that gravity is bent nearby our Sun \cite{Anania:2005dy}, of course if the real explanation is not something more conventional, like the thermal radiation pressure \cite{Turyshev:2011yi,Bertolami:2008qb}.

At the end we would like to compare our findings with the existing similar  results in the literature, e.g. \cite{Scharnhorst:1998ix,Drummond:1979pp}.
In \cite{Scharnhorst:1998ix}, using the Casimir effect, the authors showed that vacuum polarization effects may lead to superluminal propagation of photons in between the plates (since the vacuum polarization effects are suppressed there). While this is a flat space result, it is indicative that superluminality may arise in completely physical setups.
 In \cite{Drummond:1979pp}  it was argued that the quantum corrections in curved space-time are able to introduce tidal gravitational forces on the photons which in general alter the characteristics of propagation, so that in some cases photons travel at speeds greater than unity. In that case it is actually the low-frequency limit of the phase velocity that is superluminal. This indicates that propagation of quantum fields in curved space-time is a very non-trivial problem, and surprising results may be derived. It should be noted that superluminality does not always lead to paradoxes, since in both of the above mentioned cases it is impossible to send signals backward in time. While work presented in \cite{Scharnhorst:1998ix,Drummond:1979pp} is perturbative, our analysis is exact since it based on the exact solution of the Green's function in curved space-time.
It is interesting that our analysis also indicate that lower frequency modes propagate faster than high frequency modes, in good agreement with \cite{Drummond:1979pp}.

Finally, we emphasis again that we found only the phase velocity to be superluminal. If the group velocity is not superluminal, then the Green's function does not have support outside the light cone, and causality is preserved. Strictly speaking, a second order linear wave equation can not have a "wavefront" propagating faster than the speed of light. However, this statement does not affect velocity of an individual frequency component of the phase. It is only when one takes into account all the frequencies (where the higher frequencies give the dominant contribution) that he has to obey that statement.

\begin{acknowledgments}
This work was partially supported by the US National Science Foundation, under Grants No. PHY-0914893, PHY-1066278, and  by Shanghai Institutions of Higher Learning, the Science and Technology Commission of Shanghai Municipality (11DZ2260700). We thank Y.Z. Chu, J. Wang, V. Frolov and A. Zelnikov for very useful discussions.
\end{acknowledgments}


\begin{thebibliography}{99}

\bibitem{Scharnhorst:1998ix}
  K.~Scharnhorst,
  Annalen Phys.\  {\bf 7}, 700 (1998)  [hep-th/9810221].  


\bibitem{Drummond:1979pp}
  I.~T.~Drummond and S.~J.~Hathrell,
  Phys.\ Rev.\ D {\bf 22}, 343 (1980).  

\bibitem{Hollowood:2010xh}
 T.~J.~Hollowood and G.~M.~Shore., arXiv:1006.1238 [hep-th]


\bibitem{Hollowood:2010bd}
 T.~J.~Hollowood and G.~M.~Shore., Phys.\ Lett.\ B {\bf 691}, 279 (2010) 




\bibitem{Hollowood:2008kq}
 T.~J.~Hollowood and G.~M.~Shore., JHEP {\bf 0812}, 091 (2008) 


\bibitem{Hollowood:2007ku}
T.~J.~Hollowood and G.~M.~Shore., Nucl.\ Phys.\ B {\bf 795}, 138 (2008) 


\bibitem{Hollowood:2007kt}
T.~J.~Hollowood and G.~M.~Shore., Phys.\ Lett.\ B {\bf 655}, 67 (2007) 



  \bibitem{Poisson:2003nc}
  E.~Poisson,
  Living Rev.\ Rel.\  {\bf 7}, 6 (2004)
  [gr-qc/0306052].

\bibitem{Frolov:2012jj}
  V.~Frolov and A.~Zelnikov,
  arXiv:1202.0250 [hep-th].  


\bibitem{Frolov:2003mc}
  V.~P.~Frolov, M.~Snajdr and D.~Stojkovic,
  Phys.\ Rev.\ D {\bf 68}, 044002 (2003)  [gr-qc/0304083].  


\bibitem{Chu:2011ip}
  Y.~-Z.~Chu and G.~D.~Starkman,
  Phys.\ Rev.\ D {\bf 84}, 124020 (2011)  [arXiv:1108.1825 [astro-ph.CO]].  
  
\bibitem{Chu:2012kz}
  Y.~-Z.~Chu and M.~Trodden,
  arXiv:1210.6651 [astro-ph.CO].  


\bibitem{Wiseman:2000rm}
  A.~G.~Wiseman,
 Phys.\ Rev.\ D {\bf 61}, 084014 (2000)  [gr-qc/0001025].  

\bibitem{Burko:2002ge}
  L.~M.~Burko, A.~I.~Harte and E.~Poisson,
  Phys.\ Rev.\ D {\bf 65}, 124006 (2002)

\bibitem{Bird:1975we}
  J.~M.~Bird and W.~G.~Dixon,
  Annals Phys.\  {\bf 94} (1975) 320.
\bibitem{Ehlers:1976ji}
  J.~Ehlers, A.~Rosenblum, J.~N.~Goldberg and P.~Havas,
  Astrophys.\ J.\  {\bf 208} (1976) L77.

\bibitem{Stojkovic:2005zq}
  D.~Stojkovic, F.~C.~Adams and G.~D.~Starkman,
   Int.\ J.\ Mod.\ Phys.\ D {\bf 14}, 2293 (2005)  [gr-qc/0604072].  




\bibitem{Hassani}
S. Hassani, Mathematical Physics, (Springer-Verlag, New York, 1998).

\bibitem{Cardoso:2002pa}
  V.~Cardoso, O.~J.~C.~Dias and J.~P.~S.~Lemos,
  Phys.\ Rev.\ D {\bf 67}, 064026 (2003)  [hep-th/0212168].  


\bibitem{Nemiroff:2005az}
  R.~J.~Nemiroff,
  Astrophys.\ J.\  {\bf 628}, 1081 (2005)
  [arXiv:astro-ph/0502360].

\bibitem{Wucknitz}
O.~Wucknitz,
``Talk given at the ANGLES workshop 2005 in Crete'',
http://www.astro.uni-bonn.de/~wucknitz/publications/






\bibitem{Nieto:2003rq}
  M.~M.~Nieto and S.~G.~Turyshev,
  Class.\ Quant.\ Grav.\  {\bf 21}, 4005 (2004)
  [arXiv:gr-qc/0308017].

\bibitem{Anania:2005dy}
  R.~Anania and M.~Makoid,
  arXiv:astro-ph/0502582.

\bibitem{Bertolami:2008qb}
  O.~Bertolami, F.~Francisco, P.~J.~S.~Gil and J.~Paramos,
  Phys.\ Rev.\  D {\bf 78}, 103001 (2008)
  [arXiv:0807.0041 [physics.space-ph]].


\bibitem{Turyshev:2011yi}
  S.~G.~Turyshev, V.~T.~Toth, J.~Ellis and C.~B.~Markwardt,
  Phys.\ Rev.\ Lett.\  {\bf 107}, 081103 (2011)
  [arXiv:1107.2886 [gr-qc]].
\end{thebibliography}
\end{document}